\documentclass[12pt]{article}
\usepackage{epsfig}
\usepackage{url}
\makeatletter
\def\alt{\mathrel{\mathpalette\gl@align<}}
\def\agt{\mathrel{\mathpalette\gl@align>}}
\def\gl@align#1#2{\lower.6ex\vbox{\baselineskip\z@skip\lineskip\z@
\ialign{$\m@th#1\hfil##\hfil$\crcr#2\crcr\sim\crcr}}}
\makeatother

\begin{document}
\begin{flushright}
MIFPA-10-31\\
July, 2010
\end{flushright}
\vspace*{1.0cm}
\begin{center}
\baselineskip 20pt
{\Large\bf
CP Violating Lepton Asymmetry from $B$ Decays
in Supersymmetric Grand Unified Theories
} \vspace{1cm}

{\large
Bhaskar Dutta$^*$, Yukihiro Mimura$^\#$ and Yudi Santoso$^\dagger$}
\vspace{.5cm}

$^*${\it
Department of Physics, Texas A\&M University,
College Station, TX 77843-4242, USA
}\\
$^\#${\it
Department of Physics, National Taiwan University, Taipei,
Taiwan 10617, R.O.C.
}\\
$^\dagger${\it
Department of Physics and Astronomy, University of Kansas,\\
Lawrence, KS 66045-7582, USA
}\\

\vspace{1.5cm}
{\bf Abstract}
\end{center}
We investigate the effect of the dimuon CP asymmetry from the $B$ decay modes, recently  observed at 3.2 $\sigma$ deviation from the Standard Model (SM) by the D0
collaboration, in the context of SU(5) and SO(10) GUT models.
We exhibit that  a large amount of flavor violation between the second
and the third generation is generated due to the large neutrino atmospheric mixing angle
and this flavor violation can be responsible for the observed large CP asymmetry due to the presence of new phases (not present in the CKM matrix) in the Yukawa couplings. We also study
the implication of the parameter space in these GUT models with large CP violating lepton asymmetry for different
phenomenologies, e.g., Br($\tau\rightarrow\mu\gamma$),
Br($B_s\rightarrow\mu\mu$) at the Fermilab, direct detection of
dark matter (DM) in the  ongoing detectors and measurement of muon flux from solar neutrinos at the IceCube experiment.

\thispagestyle{empty}

\bigskip
\newpage

\addtocounter{page}{-1}

\section{Introduction}
\baselineskip 18pt

Recently, the D0 Collaboration
has announced the appearance of a like-sign
dimuon charge asymmetry in the semileptonic $b$-hadron decays measurement:
$A_{sl}^b = - 0.00957 \pm 0.00251 ({\rm stat}) \pm
0.00146 ({\rm syst})$ \cite{Abazov:2010hv}.
In the Standard Model (SM),
the prediction for the asymmetry is
$A_{sl}^b ({\rm SM}) = (-2.3^{+0.5}_{-0.6})\times 10^{-4}$,
and the D0 experimental result differs from this by 3.2 standard deviation.
In the absence of CP violation this quantity clearly vanishes,
hence the D0 result leads us to a new physics (NP) which induces
some CP violating flavor changing interactions beyond the SM.

The $B\bar B$ mesons created in $p\bar p$ collisions
include both $B_d (d\bar b)$ and $B_s (s\bar b)$.
The quantities of the $B_d$ system are well measured by $B$-factories,
and the unitarity triangle seems to be closed:
$V_{ud} V_{ub}^* + V_{cd} V_{cb}^* + V_{td} V_{tb}^* = 0$.
In that case, the asymmetry from the $B_d$-$\bar B_d$ oscillation
is expected to be tiny
and a new CP violating phase should show up
in the $B_s$ system instead, namely in the $b$-$s$ transition.
The D0 and CDF Collaborations have
reported the existence of a CP violating phase, $\phi_s$, in the $B_s$ system
from the $B_s \to J/\psi \phi$ decay \cite{Aaltonen:2007he,CDF}.
In fact, their results differ from the SM prediction in a direction which is consistent with the
signature of the like-sign dimuon asymmetry reported by the D0.
This provides us an encouraging guide, pointing towards a source for
new CP violation in the $b$-$s$ transitions.

Supersymmetry (SUSY) is a very attractive candidate to build NP models.
As it is well known, SUSY models have a natural dark matter candidate which is a neutralino as the lightest
SUSY particle (LSP).
Besides, the gauge hierarchy problem can be solved and a natural aspect of
the theory  can be developed from the weak scale to the ultra high
energy scale.
In fact, the gauge coupling constants of the Standard Model gauge symmetries
can unify at a high scale using the  renormalization group equations
(RGEs)
of the minimal SUSY standard
model (MSSM). This indicates the existence of a grand unified
theory (GUT) as the underlying principle for physics at the very high energy.
 The well motivated SUSY GUT models
have always been subjected to intense experimental and theoretical investigations.
Identifying a GUT model, as currently is, will be a major focus of the upcoming experiments.

In SUSY models, the SUSY breaking mass terms for the squarks and sleptons
must be introduced, and they provide sources for
flavor changing neutral currents (FCNCs) and CP violation beyond the
Kobayashi-Maskawa theory.
In general, soft breaking terms generate too large FCNCs,
hence
a flavor universality is often assumed
in the squark and slepton mass matrices
to avoid large FCNCs
in the meson mixings and lepton flavor violations (LFV) \cite{Gabbiani:1988rb}.
The flavor universality is expected to be realized by the Planck scale physics.
However,
even if the universality is realized at a high energy scale such as the GUT scale
or the Planck scale,
non-universality in the SUSY breaking sfermion masses is still generated
through the evolutions of the RGEs,
and this can lead to some small flavor violating transitions
which could possibly be observed in the ongoing experiments.
In some MSSM models
with right-handed neutrinos,
the induced FCNCs from the RGE effects are not large in the quark sector,
while sizable effects can be generated in the lepton sector due to
the large neutrino mixing angles \cite{Borzumati:1986qx}.
Within GUTs, however, loop effects due to the large neutrino mixings
can induce sizable FCNCs also in the quark sector
since the GUT scale particles which connect the quark and lepton sectors can propagate in the loops~\cite{Barbieri:1994pv}.
As a result, the patterns of the induced FCNCs
highly depend on the unification scenario
and the heavy particle contents.
Therefore, it is important to investigate
FCNC effects to obtain a footprint of the GUT physics.

If the quark-lepton unification is manifested
in a GUT model,
the flavor violation in the $b$-$s$ transition
can be responsible for the large atmospheric neutrino mixing \cite{Moroi:2000tk},
and
thus, the amount of flavor violation in the $b$-$s$ transition
(the second and the third generation mixing),
which is related to the $B_s$-$\bar B_s$ mixing and its phase,
 has to be related to the $\tau \to \mu\gamma$ decay
\cite{Dutta:2006gq,Parry:2005fp,Dutta:2008xg,Hisano:2008df,Dutta:2009iy,Parry:2010ce}
for a given particle spectrum.
The branching ratio of the $\tau \to \mu\gamma$ decay
is being measured at the $B$-factory,
and thus, the future results on LFV and from the ongoing measurement of the
$B_s$-$\bar B_s$ mixing phase
will provide important information to probe
the GUT scale physics.

In Refs.\cite{Dutta:2006gq,Dutta:2008xg,Dutta:2009iy},
the authors studied the correlation between
the branching ratio of $\tau\to\mu\gamma$ and
the phase of the $b$-$s$ transition
in the frameworks of SU(5) and SO(10) GUT models.
While performing the analysis, it is important to pay attention to
 the dependence on $\tan\beta$,
which is the ratio of the vacuum expectation values of
up- and down-type Higgs bosons.
In the case of $\tan\beta \alt 20$,
the gluino box diagram can dominate the SUSY contribution
to the $B_s$-$\bar B_s$ mixing amplitude,
while for large $\tan\beta \agt 30$
it can be dominated by the double
penguin diagram contribution \cite{Hamzaoui:1998nu,Buras:2001mb,Foster:2004vp}.
When the Dirac neutrino Yukawa coupling is the
origin of the FCNC (we call this case as minimal type of SU(5)),
the $\tau\to\mu\gamma$ constraint gives
a strong bound on the phase of the $B_s$-$\bar B_s$ mixing
for smaller $\tan\beta$.
On the other hand,
when the Majorana neutrino Yukawa coupling
is the source for the FCNC, both left- and right-handed
squark mass matrices can have off-diagonal elements
(we call this case as minimal type of SO(10)),
the gluino box contribution is enhanced and larger
$B_s$-$\bar B_s$ phase is possible compared to the SU(5) case.
The double penguin contribution is proportional to
$\tan^4\beta$,
while the Br($\tau\to\mu\gamma$) is proportional to
$\tan^2\beta$.
As a result, for both SU(5) and SO(10) cases,
a large phase of $b$-$s$ transition is allowed.
In that case, however, Br($B_s\to\mu\mu$) constraint
is more important
since it is proportional to
$\tan^6\beta$ \cite{Choudhury:1998ze}.
In other words, when the phase of the $b$-$s$ transition
is large due to the double penguin contribution,
Br($B_s\to\mu\mu$) has to be also enhanced.
In fact, in \cite{Dutta:2009iy}
it was shown that a lower bound from Br($B_s\to\mu\mu$)
is obtained in the case of SU(5) GUT.

In Ref.\cite{Dutta:2009hj},
we also investigated the implication on the dark matter detection
from the large $B_s$-$\bar B_s$ mixing.
Assuming that the neutralino LSP is the dark matter candidate,
the SUSY parameters are restricted by the relic density constraint.
It was shown that the funnel region, in which the relic density constraint is
satisfied through annihilation near heavy Higgs pole, is favored by the flavor
solution. Moreover, there are some correlation between flavor changing processes
and the neutralino direct detection cross section through the dependency on the
CP-odd Higgs mass, $m_A$.

In this paper,
we will sort out the GUT models,
where the FCNC is due to the atmospheric neutrino mixing,
to obtain a large CP asymmetry of the $B$ decays.
This investigation is important if the reported size of the like-sign
dimuon charge asymmetry persists in the future
with a smaller error.
We show that it prefers the SO(10) type boundary condition where
Majorana neutrino couplings induce the FCNC
and
both left- and right-handed squark mass matrices
have off-diagonal elements.
Especially, for the SU(5) boundary condition
where the Dirac neutrino Yukawa coupling
induces the FCNC to produce a large CP asymmetry,
a large value of $\tan\beta$ is required
and the SUSY mass spectrum is restricted.
We will also study the implication of the dark matter direct and indirect detections
from the constraints on the SUSY mass spectrum.

This paper is organized as follows: in section 2 we discuss
the resent results of CP violation in $B_s$ decays; in section 3,
we discuss the sources of flavor changing neutral currents (FCNC)
in the context of SUSY GUTs; in section 4, we show constraints
arising from the experimental constraints on different FCNC processes;
in section 5, we discuss the constraints from the dark matter content
of the universe and predictions related to the direct and indirect detection
experiments; and we conclude in section 6.

\section{CP violation in $B_s$ decays}

The dimuon like-sign asymmetry $A_{\rm sl}^b$ by D0
deviates from the SM prediction by 3.2 $\sigma$.
The $B\bar B$ samples created at the Tevatron
includes both $B_d$ and $B_s$,
and the asymmetry can be written as
$A_{\rm sl}^b = (0.506\pm 0.043)a_{\rm sl}^d +
(0.494\pm 0.043) a_{\rm sl}^s$ \cite{Abazov:2010hv}.
The pieces of $a_{\rm sl}^{q}$ $(q=d,s)$ can be defined as
$a_{\rm sl}^{q} = (r_q - \bar r_q)/(r_q+\bar r_q)$
where $r_q$ and $\bar r_q$ are the
ratios of $B$-$\bar B$ mixings:
$r_q= P(\bar B\to B)/P(\bar B\to \bar B)$
and
$\bar r_q= P(B\to \bar B)/P(B\to B)$.
Since the $B_d$ system is consistent with experiments,
we assume that the CP asymmetry in the $B_d$-$\bar B_d$ mixing
is negligible.
When we take
the experimental data on dimuon asymmetry by CDF (1.6 fb$^{-1}$)  and  on ``wrong-charge" asymmetry in the semileptonic $B_s$ decay by D0 into account,
the asymmetry in $B_s$-$\bar B_s$ mixing is extracted as follows\cite{Dobrescu:2010rh}
\begin{equation}
a_{\rm sl}^s = (-12.7 \pm 5.0) \times 10^{-3},
\end{equation}
which deviates from the SM prediction by about 2.5 $\sigma$.

When $\Gamma_{12}^s \ll M_{12}^s$ ($M_{12}^s$ is the mixing amplitude
and $\Gamma_{12}^s$ is the absorptive part of the mixing),
$a_{\rm sl}^s$ is given as \cite{Hagelin:1981zk}
\begin{equation}
a_{\rm sl}^s = {\rm Im}\, \frac{\Gamma_{12}^s}{M_{12}^s}
 =  \left|\frac{\Gamma_{12}^s}{M_{12}^s}\right| \sin\phi_s,
\end{equation}
where $\phi_s$ is the argument of $\Gamma_{12}^s/M_{12}^s$.

In many  NP models, the
$\Delta B =2$ ($B$ is beauty) Hamiltonian
can be easily modified 
(e.g.
see for recent works motivated by the D0 results \cite{Dobrescu:2010rh,recent}).
We parameterize the $M_{12}^s$ as
\begin{equation}
M_{12}^s = M_{12,\rm SM}^s + M_{12,\rm NP}^s
=C_s \, M_{12,\rm SM}^s \, e^{2i \phi_{B_s}},
\end{equation}
where $C_s$ is a real positive number. From
the measurement of the mass difference, $\Delta M_s = 2|M_{12}^s|$,
the experimental result is consistent with $C_s =1$.
When the $\Delta B =1$
Hamiltonian is  same
(allowing modification at 10\% level)
as the SM,
even in the presence of new physics,
the phase of $\Gamma_{12}^s$ is almost same as that of SM,
which is tiny $\sim 0.04$
(in usual phase convention where $V_{cb} V_{cs}^*$ is almost real).
In that case, $\phi_s$ is the same as the phase
($-2\beta_s = -2(\beta_s^{\rm SM}+ \phi_{B_s})$) measured by
$B_s\to J/\psi \phi$ decay observation.
Using the $B_s\to J/\psi \phi$ decay, the decay width difference
$\Delta \Gamma_s = 2 |\Gamma_{12}^s| \cos \phi_s$ is also measured.
The parameters of $B_s$-$\bar B$ oscillations,
$\beta_s$ and $\Delta \Gamma_s$, have been measured
at the Tevatron \cite{Aaltonen:2007he},
and the CDF Collaboration showed their recent analysis
till 5.2 fb$^{-1}$ of data \cite{CDF}.
It appears that the data statistics is very different over the periods ($0-2.8$ fb$^{-1}$ and $2.8-5.2$ fb$^{-1}$).
We will adopt their analysis for $0-5.2$ fb$^{-1}$.
The CDF result on the phase $2\beta_s$ differs from the SM prediction $2\beta_s^{\rm SM} \sim 0.04$
at 1 $\sigma$ level (D0 shows about 2 $\sigma$ deviation for
the same measurement for 2.8 fb$^{-1}$ data~\cite{Aaltonen:2007he}),
and the signature of the phase is consistent with the sign required to explain
the anomalous like-sign dimuon charge asymmetry by D0.

The SM prediction on $\Gamma_{12}^s/M_{12}^s$ is given by
Lenz-Nierste \cite{Lenz:2006hd}
\begin{equation}
\left|\frac{\Gamma_{12}^s}{M_{12}^s}\right|_{\rm SM} =
(4.97 \pm 0.94) \times 10^{-3}.
\label{Lenz-Nierste}
\end{equation}
It was pointed out that theoretical prediction
of the absolute value of $a_{\rm sl}^s$ is bounded from above
and
the bound is a little bit too small
to explain the dimuon asymmetry by D0 if $\Gamma_{12}^s$ is not modified
\cite{Dobrescu:2010rh,Hou:2007ps}.
This is because the experimental measurement of
$\Delta M_s = 2|M_{12}^s|$ is consistent with the SM prediction (which means
$C_s \simeq 1$). 
Using the simple relation:
\begin{equation}
 \left(\frac{\Delta \Gamma_s}{\Delta M_s}\right)^2
  + (a_{\rm sl}^s)^2
= \left|\frac{\Gamma_{12}^s}{M_{12}^s} \right|^2
= \frac{1}{C_s^2} \left|\frac{\Gamma_{12}^s}{M_{12}^s} \right|^2_{\rm SM},
\label{parameter}
\end{equation}
we illustrate the current situation in the Figure 1.
The solid circles correspond to the 1 $\sigma$ boundaries
in the case of $C_s = 1$ by using the
SM prediction by Lenz-Nierste.
The colored (blue and red) solid curves correspond to the
measurement from the $B_s \to J/\psi \phi$ decay by CDF ($0-5.2$ fb$^{-1}$).
We assume that $\phi_s = -2\phi_{B_s}$, which means that
${\rm arg}\,\Gamma_{12}^s = {\rm arg}\,\Gamma_{12,\rm SM}^s$.
The horizontal colored (yellow) band is the 1 $\sigma$ region of $a_{\rm sl}^s$.
As one can see, the 1 $\sigma$ regions do not match well,
independent of the choice of $\phi_s$,
where $\tan\phi_s = a_{\rm sl}^s/(\Delta \Gamma_s/\Delta M_s)$.

For the current situation, the combined analysis
has a large error for $a_{\rm sl}^s$,
the discrepancy is not very serious if the phase $\phi_s$ is $O(1)$ rad.
However, the dimuon asymmetry measured by the D0 alone has a large center value
(which corresponds to $a_{\rm sl}^s = (-19.4\pm 6.1)\times 10^{-3}$),
and if the measurement of the dimuon asymmetry
(or the ``wrong-charge" muon asymmetry in the semileptonic $B_s$ decays)
becomes accurate in the future keeping the large center value of $a_{\rm sl}^s$,
one has to resolve this issue.
To implement such possible future constraints,
one can consider the following three typical remedies.

\begin{figure}[tbp]
 \center
 \includegraphics[viewport = 10 10 230 220,width=8cm]{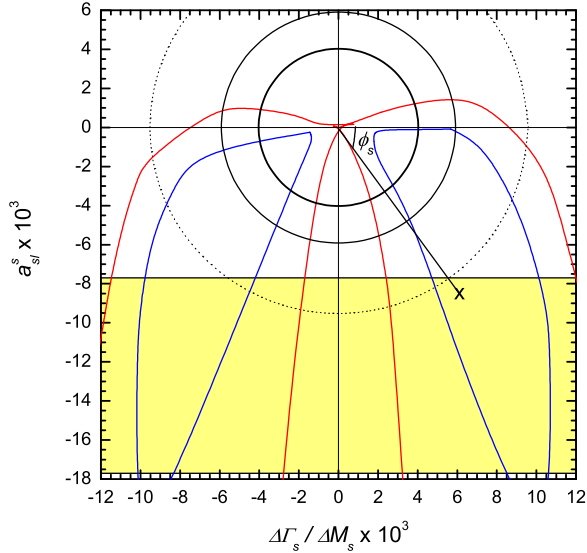}
 \caption{
The experimental and theoretical regions in
the $a_{\rm sl}^s$-$\Delta \Gamma_s/\Delta M_s$ plane.
The yellow region is the 1 $\sigma$ region of
the combined data from the semi-leptonic $B$ decays.
The red and blue lines are 95\% and 68\% boundaries
from the $B_s\to J/\psi \phi$ decay, assuming that
the phase of $\Gamma_{12}^s$ is the same as the phase of
$\Gamma_{12,\rm SM}^s$.
The solid circles are the theoretical 1 $\sigma$ boundaries
using the numerical calculation in Ref.\cite{Lenz:2006hd}.
The dotted circle corresponds to the conservative theoretical region
when one implements the remedies 2 and 3 in the text.
}
\end{figure}

\begin{enumerate}

\item
Add a $\Delta B =1$ effective
Hamiltonian to modify $\Gamma_{12}^s$ or $\Gamma_{12}^d$
\cite{Dighe:2010nj,Bauer:2010dg,Deshpande:2010hy,Bai:2010kf}.

This is a direct resolution of this issue.
If in the future measurements the phase of $B_s \to J/\psi \phi$ is
really diminished, this type of resolution will be needed.
(In fact, the recent CDF data for the period $2.8-5.2$ fb$^{-1}$
may indicate that the $B_s\to J/\psi \phi$ phase is almost zero.)

\item
Non-perturbative effects \cite{Hou:2007ps,Deshpande:2010hy}.

In this case, the numerical number in Eq.(\ref{Lenz-Nierste})
is obtained by a two parameter
expansion, $\Lambda_{\rm QCD}/m_b$ and $\alpha_s (m_b)$,
using operator product expansion
and heavy quark expansion.
In fact, it is known that non-perturbative effects may dominate
in the $D^0$-$\bar D^0$ meson mixings,
and it may be true that there is a large long distance contribution
in the $\Gamma_{12}^s$ (e.g. the intermediate states include $D_s^+ D_s^-$,
which may lead to large non-perturbative effects).
In the case of $B_s$-$\bar B_s$ mixings, each term of the
next to leading order is about 30\% of the leading order,
and the expansion may be more reliable
than in the case of $D$-$\bar D$ mixing.
However, a careful treatment is needed since the series may not be converging.
Actually, the term which has the largest uncertainty in the next to leading order
calculation gives
a negative contribution to $\Gamma_{12}^s$,
and the true numerical value may be larger than in Eq.(\ref{Lenz-Nierste}).
In that sense, the discrepancy is not so serious
unless the $B_s \to J/\psi\phi$ phase will become tiny with a small error
in future.

\item
Use the uncertainty of the Bag parameter $B_{B_s}$
and the decay constants $f_{B_s}$.

The mixing amplitudes are proportional to $B_{B_s} f_{B_s}^2$,
which has about 40\% error.
This factor is canceled in the ratio of $\Gamma_{12}^s/M_{12}^s$,
and the SM prediction in Eq.(\ref{Lenz-Nierste}) does not have the ambiguity.
The parameter $C_s$ in Eq.(\ref{parameter}) may have the 40\% error.
However, since the ratio of the hadronic quantities for $B_d$ and $B_s$,
related to the SU(3) flavor violation,
is more accurate \cite{Aubin:2009yh}
\begin{equation}
\xi = \frac{f_{B_s}\sqrt{B_{B_s}}}{f_{B_d}\sqrt{B_{B_d}}}= 1.23\pm 0.04,
\end{equation}
the mass difference of $B_d$ restrict the uncertainty of $C_s$
less than 10\%
because of the relation
\begin{equation}
\left|\frac{M_{12}^s}{M_{12}^d}\right|_{\rm SM} =
\frac{M_{B_s}}{M_{B_d}} \left|\frac{V_{ts}}{V_{td}}\right|^2 \xi^2.
\end{equation}
Therefore,
if one uses the full uncertainty of
$B_{B_s} f_{B_s}^2$,
one also has to modify $|M_{12}^d|$ in the same rate as of $|M_{12}^s|$.
In general, it is possible to do that, but one should be careful about the
argument of $M_{12}^d$ since the $\sin2\beta$ measurement
is consistent with the SM.

In SUSY models, the
argument of $M_{12}^d$ is related to the 13 mixing/23 mixing of the
squark mass matrices, and in  models where the FCNC is induced by
the Dirac/Majorana neutrino Yukawa couplings, it is related to the
size of the 13 neutrino mixing.

\end{enumerate}

In this paper, we consider the case where
$\Gamma_{12}^s \simeq \Gamma_{12,\rm SM}^s$
and the phase $\phi_s$ comes from the $M_{12}^s$ phase
$\phi_s = -2\phi_{B_s}$, in SUSY models
with $R$-parity conservation.

The dotted circle in the Figure 1 corresponds
to the region when we used the 2 $\sigma$ error of the
Eq.(\ref{Lenz-Nierste}) and
40\% error from the $B_{B_s} f_{B_s}^2$
and $\Gamma_{12}^s = \Gamma_{12,\rm SM}^s$.
The absolute value of the phase $\phi_s$ should be large
$\sim 50^{\rm o} - 70^{\rm o}$
to explain the large CP asymmetry $a_{\rm sl}^s$.

By definition in Eq.(\ref{parameter}), we obtain
\begin{equation}
\sin^2\phi_{B_s} =
\frac{\left(\frac{A^{\rm NP}_s}{A^{\rm SM}_s}\right)^2-(1-C_s)^2}{4C_s},
\label{relation-phiBs}
\end{equation}
where $A_s^{\rm NP} = | M_{12,\rm NP}^s |$ and
$A_s^{\rm SM} = | M_{12,\rm SM}^s |$.
When $C_s \simeq 1$, we obtain $2\sin\phi_{B_s} \simeq A^{\rm NP}_s/A^{\rm SM}_s$.
Therefore, the NP contribution of $M_{12}^s$
should be comparable to the SM contribution
to obtain the large phase $\phi_s \simeq -2\phi_{B_s}$.

In SUSY models
(for an early study of the dimuon asymmetry
in SUSY models, see \cite{Randall:1998te}), we need to realize
 $A_s^{\rm NP} \sim A_s^{\rm SM}$
in order to explain the current combined data of CP asymmetry in $B$ decay.
As it is  mentioned already that in GUT models,
the Dirac/Majorana neutrino Yukawa coupling
can be a source for FCNC  even in the quark sector.
When the quark-lepton unification is manifested,
the amount of $A_s^{\rm NP}$ is related to the
lepton flavor violation $\tau \to \mu\gamma$,
and will be bounded by the constraint on
Br($\tau \to \mu\gamma$).
The main purpose of this paper is
to study how to obtain a large $A_s^{\rm NP}$
when there is quark-lepton unification and after satisfying all the other experimental  constraints.

\section{FCNC sources in SUSY GUTs}

In SUSY GUT theories,
it is often assumed that
the SUSY breaking sfermion masses
are flavor-universal,
but the off-diagonal elements of the mass matrices
 are generated by the loop effects.
The FCNC sources are the Dirac/Majorana neutrino Yukawa couplings,
which are responsible for the large neutrino mixings
\cite{Borzumati:1986qx,Barbieri:1994pv}.
Since the left-handed leptons $(L)$ and
the right-handed down-type quarks $(D^c)$
are unified in $\bar{\bf 5}$,
the Dirac neutrino Yukawa couplings can be written as
$Y_\nu{}_{ij} \bar{\bf 5}_i N^c_j H_{\bf 5}$,
where $N^c$ is the right-handed neutrino.
The flavor non-universality of the SUSY breaking $\tilde D^c$
masses is generated
by the colored Higgs and the $N^c$ loop diagram \cite{Moroi:2000tk},
and the non-universal part of the mass matrix is
$\delta M_{\tilde D^c}^2 \simeq
- \frac1{8\pi^2} (3m_0^2+A_0^2) Y_\nu Y_\nu^\dagger \ln(M_*/M_{H_C})$,
where $M_*$ is a cut-off scale (e.g. the Planck scale), $M_{H_C}$ is
a colored Higgs mass,
$m_0$ is the universal scalar mass and
$A_0$ is the universal scalar trilinear coupling.
The left-handed Majorana neutrino coupling $LL\Delta_L$
($\Delta_L$ is an SU(2)$_L$ triplet)
can also provide  contributions to the light neutrino mass
(type II seesaw \cite{Schechter:1980gr}),
and can generate the FCNC in the sfermion masses
when the fermions are unified.

As a convention in this paper,
we will call the model with the FCNC source arising from
the Dirac neutrino Yukawa coupling as the minimal type of SU(5).
In this case, the off-diagonal elements of $\bf 10$
($Q,U^c,E^c$) representations are small because they originate from
the CKM mixings.
In a competitive model which we call the minimal type of SO(10),
the Majorana couplings, which contribute to the neutrino mass,
generate the off-diagonal elements for all sfermion species
since the Majorana couplings $f_{ij} L_iL_j\Delta_L$ can be unified to the
$f_{ij}{\bf 16}_i\ {\bf 16}_j\ \overline{\bf 126}$ coupling \cite{Babu:1992ia}.

We note that
when the source is not specified, such as the case for Dirac Yukawa coupling,
the off-diagonal elements (of $\bf 10$ multiplets in SU(5)) can be large in general.
In our convention of SU(5) and SO(10) models,
the source of the off-diagonal elements are specified,
and only the right-handed down-type squark
mass matrix has sizable off-diagonal elements in SU(5),
while both left- and right-handed squark mass matrices can have
sizable off-diagonal elements in SO(10).

The non-universal part generated
from the Dirac/Majorana couplings, $Y_\nu$ and $f$,
is proportional to $Y_\nu Y_\nu^\dagger$ and $f f^\dagger$.
In general, therefore,
the squark and slepton mass matrices due to the loop correction can be
parameterized as
\begin{equation}
M_{\tilde F}^2 = m_0^2 [{\bf 1} - \kappa_F U_F {\rm diag} (k_1,k_2,1) U_F^\dagger],
\end{equation}
where $F= Q,U^c,D^c,L,E^c$.
The unitary matrices $U_F$
is equal to the neutrino mixing matrix in a limit \cite{Dutta:2006zt}.
The quantity $\kappa_F$ denotes the amount of the off-diagonal elements
and it depends on the sfermion species.
In the minimal type of SU(5),
since the off-diagonal elements of the SUSY breaking mass matrix
for the left-handed lepton doublet get the correction,
$\delta M_{\tilde L}^2 \simeq -1/(8\pi^2) (3m_0^2+A_0^2) \sum_k
(Y_\nu)_{ik} (Y_\nu)_{jk} \ln (M_*/M_k)$ where $M_*$ is a scale
that the flavor universality is realized,
$\kappa_L$ can be written as $\kappa_L \sim (Y_\nu^{\rm diag})_{33}^2
(3+A_0^2/m_0^2)/(8\pi^2)\ln (M_*/M_R)$, where $M_R$ is a Majorana mass
of the right-handed neutrino.
If the GUT threshold effects are neglected, we have
$\kappa_{\bar {\bf5}} = \kappa_L = \kappa_{D^c}$,
and $U_{\bar{\bf 5}} = U_L = U_{D^c}$.
In general, the fermion mass matrices arise from the sum of the Yukawa terms and
the equality of $U_{D^c}$ and $U_L$ can be completely broken
when there are cancellations among the minimal Yukawa term and
additional Yukawa terms.
Here, we consider a model
where the (near) equality between $U_{D^c}$ and $U_L$ (especially for
23 mixing angle of them) is maintained
as a ``minimal type" assumption.
The assumption is natural if there is a dominant Yukawa contribution
and corrections to fit realistic masses and mixings are small.
The unitary matrices for
$Q$, $U^c$, $E^c$ species are related to the CKM matrix,
and can generate only negligible effects to the following discussion.
Therefore, for the SU(5) boundary condition, we assume, is as follows:
\begin{equation}
{\rm SU(5)} : \quad \kappa_L = \kappa_{D^c}, \quad U_L = U_{D^c}, \quad \kappa_Q = \kappa_{U^c}= \kappa_{E^c}=0.
\label{SU(5)}
\end{equation}
This boundary condition will be used for  discussions in the following sections.
In the minimal type of SO(10) model, all $U_F$ can have large mixings
responsible for the neutrino mixings.
If the threshold effects are neglected, one finds
$\kappa_{\bf 16} \simeq 15/4 (f_{33}^{\rm diag})^2 (3+A_0^2/m_0^2)/(8\pi^2)
\ln M_*/M_U$, where $M_U$ is a SO(10) unification scale.
In general, however, the equality of all $\kappa_F$ in the SO(10) boundary condition
is broken by  threshold effects.
A detail physical interpretation of this parameterization is given in
\cite{Dutta:2009iy,Dutta:2006zt}.
The SO(10) boundary condition,
we assume, is as follows:
\begin{equation}
{\rm SO(10)}: \quad \kappa_Q = \kappa_{U^c}= \kappa_{D^c}=\kappa_{L} = \kappa_{E^c},
 \quad U_Q = U_{U^c}=U_{D^c}=U_{L}=U_{E^c}.
\label{SO(10)}
\end{equation}

When the Dirac neutrino Yukawa coupling $Y_\nu$ or the
Majorana coupling $f$ is hierarchical,
we obtain $k_1,k_2 \ll 1$ and
then the 23 element of the sfermion mass matrix
is $-1/2 m_0^2 \kappa \sin 2\theta_{23} e^{i \alpha}$.
The magnitude of the FCNC between 2nd and 3rd generations
is controlled by $\kappa \sin2\theta_{23}$,
where $\theta_{23}$ is the mixing angle in the unitary matrix.
The phase parameter $\alpha$ also originates from the unitary matrix,
and it will be the origin of a phase of the FCNC contribution.

As it is mentioned that
it is preferable to modify the absolute value of $B_d$-$\bar B_d$
mixing amplitude $|M_{12}^d|$ without modifying its argument
in order to enhance the asymmetry $a_{\rm sl}^s$.
The 13 element of the sfermion mass matrix is
$\kappa (-1/2 k_2 \sin2\theta_{12}\sin\theta_{23}
+ e^{i\delta} \sin\theta_{13} \cos\theta_{23}$) \cite{Dutta:2006zt},
where $\theta_{ij}$ are the mixing angles and $\delta$ is a
phase in the unitary matrix.
Choosing small values for the parameters $k_2$ and $\theta_{13}$,
one can realize the preferred situation.

\section{Constraints from the FCNC processes in SUSY GUTs}

In the MSSM with flavor universality, the chargino box diagram dominates
the SUSY contribution to $M_{12}^s$.
In the general parameter space of the soft SUSY breaking terms,
the gluino box diagram can dominate the SUSY contribution for a lower $\tan\beta$
(i.e. $\tan\beta \alt 20-30$).
The gluino box contribution is enhanced if both left- and right-handed
down-type squark
mass matrices have off-diagonal elements \cite{Ball:2003se},
and
therefore,
it is expected that the SUSY contribution to the $B_s$-$\bar B_s$
mixing amplitude is large for the SO(10) model with type II seesaw,
compared to the minimal type of SU(5) model \cite{Dutta:2008xg}.

When the lepton flavor violation is correlated to the flavor violation
in the right-handed down-type squark,
the $\tau\to \mu\gamma$ decay gives us the most important constraint
to obtain the large $B_s$-$\bar B_s$ phase \cite{Dutta:2008xg,Hisano:2008df}.
Furthermore, the squark masses are raised much more compared to
the slepton masses due to the gaugino loop contribution
since the gluino is heavier compared to the Bino and the Wino at low energy (assuming the gaugino mass universality at a high energy scale such as the GUT scale),
and thus the lepton flavor violation will be more sizable compared to the quark
flavor violation.
The current experimental bound on the branching ratio of $\tau\to\mu\gamma$
is \cite{Hayasaka:2007vc}
\begin{equation}
{\rm Br}(\tau\to\mu\gamma) < 4.4 \times 10^{-8}.
\label{taumugamma}
\end{equation}

In order to allow for a large phase in the $B_s$-$\bar B_s$ mixing,
a large flavor-universal scalar mass 
at the cutoff scale
is preferable.
The reasons are as follows.
The gaugino loop effects are flavor invisible
and they enhance the diagonal elements of the scalar mass matrices
while keeping the off-diagonal elements unchanged.
If the flavor universal
scalar masses at the cutoff scale become larger,
both Br($\tau\to\mu\gamma)$ and $\phi_{B_s}$ are suppressed.
However, Br($\tau\to\mu\gamma)$ is much more suppressed compared to $\phi_{B_s}$
for a given $\kappa \sin2\theta_{23}$
because the low energy slepton masses are sensitive to $m_0$
while the squark masses are not so sensitive
due to the gluino loop contribution to their masses.
The large Higgsino mass, $\mu$, is also helpful
to suppress the dominant chargino contribution
of $\tau\to\mu\gamma$.
The subdominant neutralino contribution, however, will
become large when $\mu$ is large due to the large left-right
stau mixing.

\begin{figure}[tbp]
 \center
 \includegraphics[viewport = 0 10 290 220,width=9cm]{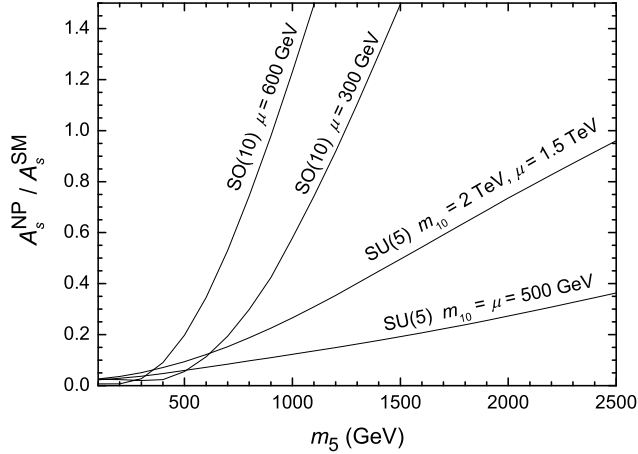}
 \caption{
The possible SUSY contributions are plotted when
the $\tau\to\mu\gamma$ bound is saturated.
The SO(10) boundary condition can give a larger
SUSY contribution than the SU(5) boundary condition.
We choose $m_{1/2} = 300$ GeV and $\tan\beta =10$ for this plot.
The detail to draw the plot is written in the text.
The relation between the CP phase $\phi_s = -2\phi_{B_s}$
and $A_s^{\rm NP}/A_s^{\rm SM}$ is given in
Eq.(\ref{relation-phiBs}).
}
\end{figure}

In figure 2,
we plot the magnitude of $A_s^{\rm NP}/A_s^{\rm SM}$
as a function of $m_5$ (the SUSY breaking mass of
$\bar{\bf 5} = (D^c,L)$ at the unification scale),
when the $\tau\to\mu\gamma$ bound, Eq.(\ref{taumugamma}),
is saturated,
for various mass parameters in the case of $\tan\beta=10$.
We choose the unified gaugino mass as $m_{1/2} = 300$ GeV,
and
the universal scalar trilinear coupling as $A_0=0$.
In the case of SU(5), the SUSY breaking mass of
${\bf 10} = (Q,U^c,E^c)$, $m_{10}$, can be
different from $m_{5}$.
As one can see,
in order to achieve $A_s^{\rm NP}/A_s^{\rm SM} \sim 1$,
the mass parameters should be around 2 TeV.
In the case of SO(10), the sfermion masses are unified,
$m_0 = m_5 = m_{10}$, and thus we only change
$\mu$ in the two plots in the figure.
As one can see,
the NP contribution in SO(10)
can be much bigger than SU(5) case.
This is the consequence of the fact
that both left- and right-handed squark
mass matrix can have sizable off-diagonal elements
from the Majorana neutrino coupling
in SO(10) case.

In the lower $\tan\beta$ case, however,
the amount of non-universality $\kappa$
has to be large $\agt 0.3$ to achieve
$A_s^{\rm NP}/A_s^{\rm SM} \sim 1$, especially in SU(5).
Such a sizable $\kappa$ is possible if $A_0/m_0$ is large,
but the large $\kappa$ is not preferable
as long as it is the RGE induced origin from the Planck scale
and the GUT scale.
Besides, the muon $g-2$ anomaly \cite{g-2} is also suppressed
when $\tau\to\mu\gamma$ is suppressed.
This is not good since the deviation of $g-2$ from the SM prediction
is now estimated about 3.2 \cite{Davier} - 4 $\sigma$ level \cite{Teubner:2010ah}.
When $\tan\beta$ is larger ($> 30-40$),
the so called double Higgs penguin diagram
dominates the contribution rather than the box diagram,
and $\kappa$ can be smaller $\alt 0.1$ to achieve
$A_s^{\rm NP}/A_s^{\rm SM} \sim 1$.
In this case, we do not need to suppress $\tau\to\mu\gamma$,
and the muon $g-2$ anomaly can be explained.

The double Higgs penguin (flavor changing neutral Higgs interaction)
is generated as follows \cite{Choudhury:1998ze}.
In SUSY models, only the holomorphic coupling is allowed
for the Yukawa coupling. When SUSY is broken,
the non-holomorphic coupling is generated by the finite corrections.
For the down-type quark, the Yukawa coupling is
\begin{equation}
{\cal L}^{\rm eff} = Y_d Q D^c H_d + Y_d^\prime Q D^c H_u^*.
\end{equation}
The second term is the non-holomorphic term,
and $Y_{d}^\prime{}_{23}$ and $Y_{d}^\prime{}_{32}$
are roughly proportional to $\tan\beta$.
Since we work on the basis where the down-type quark mass matrix
($M_d = Y_d v_d + Y_d^\prime v_u$) is diagonal,
the following flavor changing Higgs coupling can be obtained:
\begin{equation}
Y_d^\prime Q D^c H_u^* - Y_d^\prime \frac{v_u}{v_d} Q D^c H_d.
\end{equation}
The dominant flavor changing neutral Higgs
coupling is roughly obtained from the second term,
and it is proportional to $\tan^2\beta$.
The $B_s$-$\bar B_s$ mixing can be generated from a
double penguin diagram,
the mixing amplitude is proportional to $\tan^4\beta$.
Since the Br($\tau\to\mu\gamma$) is proportional to $\tan^2\beta$,
the double penguin contribution for large $\tan\beta$
is preferable to obtain $A_s^{\rm NP}/A_s^{\rm SM} \sim 1$
satisfying the $\tau\to\mu\gamma$ constraint in GUT models.

The effective flavor changing Higgs couplings are written as
\begin{equation}
X_{RL}^{Sij} (\bar d_i P_R d_j) S^0 + X_{LR}^{Sij} (\bar d_i P_L d_j) S^0,
\end{equation}
where $S^0$ represents for the neutral Higgs fields, $S = [H,h,A]$,
where $H$ and $h$ stand for heavier and lighter CP even neutral Higgs fields,
and $A$ is a CP odd neutral Higgs field (pseudo Higgs field).
The couplings are
\begin{eqnarray}
X_{RL}^{Sij} &=& Y^\prime_d{}_{ij} \frac{1}{\sqrt2 \cos\beta} [\sin(\alpha-\beta),\cos(\alpha-\beta),-i], \\
X_{LR}^{Sij} &=& Y^\prime_d{}_{ji} \frac{1}{\sqrt2 \cos\beta} [\sin(\alpha-\beta),\cos(\alpha-\beta),i],
\end{eqnarray}
where $\alpha$ is a mixing angle for $h$ and $H$.
The double penguin diagram including both left- and right-handed
Higgs penguin which is proportional to the factor
\begin{equation}
\frac{\sin^2(\alpha-\beta)}{m_{H}^2} + \frac{\cos^2(\alpha-\beta)}{m_{h}^2} + \frac{1}{m_{A}^2},
\end{equation}
and the double penguin contribution is naively proportional to
$X_{RL}^{23} X_{LR}^{23}/m_{A}^2$
\cite{Hamzaoui:1998nu,Buras:2001mb,Foster:2004vp}.
On the other hand, the double left-handed (or double right-handed)
penguin contribution $\propto (X_{LR}^{23})^2$ (or $(X_{RL}^{23})^2$)
is tiny because $\cos(\alpha-\beta) \simeq 0$ and $m_A \simeq m_H$
for $\tan\beta \gg 1$ and $m_A > m_Z$.

In the flavor universal SUSY breaking, the right-handed penguin coupling
$X_{RL}^{23}$ is tiny,
and the double penguin contribution cannot be sizable
even for a large $\tan\beta$.
However, when the right-handed mixing is generated in the SUSY GUT models,
the double penguin diagram can be sizable for large $\tan\beta$.
We note that if there is a FCNC source in the right-handed squark mass matrix,
we do not need the off-diagonal elements in the left-handed squark mass matrix
in order to generate the sizable double penguin contribution.
Therefore, even in the minimal type of SU(5) model,
the double penguin contribution can be sizable when $\tan\beta$ is large.

\begin{figure}[tbp]
 \center
 \includegraphics[viewport = 0 10 290 220,width=9cm]{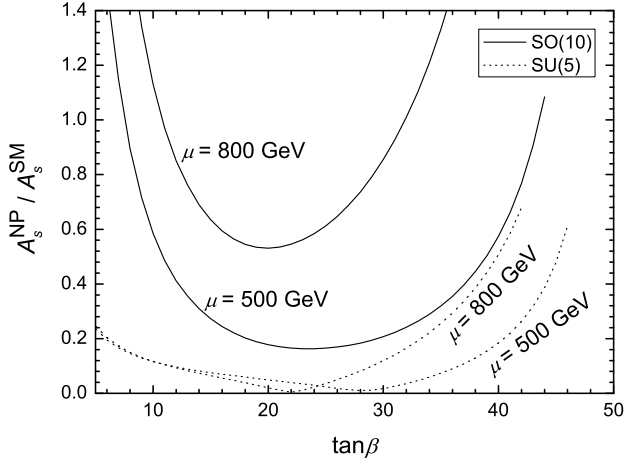}
 \caption{
The possible SUSY contributions are plotted as a function of
$\tan\beta$ when
the $\tau\to\mu\gamma$ bound is saturated.
The SO(10) boundary condition can give a larger
SUSY contribution than the SU(5) boundary condition.
We choose $m_{1/2} = 300$ GeV and $m_0 = 800$ GeV.
The detail to draw the plot is written in the text.
}
\end{figure}

In figure 3, we plot the $\tan\beta$ dependence of
$A_s^{\rm NP}/A_s^{\rm SM}$
when Br($\tau\to\mu\gamma$) saturates the experimental bound
for two Higgsino masses $\mu$.
We choose the unified gaugino mass as $m_{1/2} = 300$ GeV,
and the SUSY breaking scalar masses as $m_0 = m_5 = m_{10} = 800$ GeV, and
$A_0=0$.
We assume $m_{H_u}^2 = m_{H_d}^2$ for the SUSY breaking
Higgs mass at the unification scale just for reducing the number of parameter.
As one can see, $A_s^{\rm NP}/A_s^{\rm SM}$ becomes smaller
for $\tan\beta \sim 20$.
This is because Br($\tau\to\mu\gamma$) is proportional to $\tan^2\beta$
while
the box contribution does not depend on $\tan\beta$.
The double penguin contribution
is proportional to $\tan^4\beta$,
and thus the amplitude can become larger
for $\tan\beta> 30$.
For large $\tan\beta > 40$,
the constraint from Br($B_s\to \mu\mu$) \cite{Bsmumu}
\begin{equation}
{\rm Br}(B_s \to\mu\mu) < 4.3 \times 10^{-8},
\end{equation}
becomes more important,
since it is proportional to $\tan^6\beta$.
In the plots,
the lines are terminated when the $B_s\to\mu\mu$ bound
is saturated.

The left-handed flavor changing Higgs coupling, $X_{LR}^{23}$
can be generated by chargino diagram
even if the flavor violating source is only the CKM mixing.
Therefore,
the mixing amplitude can be enhanced
when only the right-handed down-type squark
mass matrix has the off-diagonal element such as in the SU(5) case.
The left-handed source of FCNC is also helpful
to enhance the mixing amplitude
since it can give a constructive contribution
to the left-handed penguin.
In fact, in the SO(10) boundary condition,
there is additional phase freedom from the Yukawa matrix,
and the phases of the off-diagonal elements
for left- and right-handed squark mass matrices
are independent in the basis where
the down-type quark mass matrix is real and positive
diagonal matrix.
As a consequence,
even if we fix the phase of $M_{12,\rm SUSY}^s$,
there remains one more phase freedom.
In figure 2, actually,
we choose the
additional phase in the left-handed off-diagonal element
to make the constructive contribution to the mixing amplitude.
Therefore, the mixing amplitude
under the SO(10) boundary condition
can be larger than the case in SU(5).

It is interesting to note that
the chargino contribution to $b\to s\gamma$
is destructive
when the Higgs penguin contribution is constructive
\cite{Buras:2001mb,Dutta:2009iy}.
This is roughly because the electric charge of
down quark is negative,
and the signatures of amplitudes for $b\to s\gamma$
and the finite correction of the down-type quark mass matrix
are opposite.
As a result, the SO(10) boundary condition can be
also preferable
from the $b\to s\gamma$ constraint.

We comment on the GUT threshold effects
for the boundary conditions, Eqs.(\ref{SU(5)}),(\ref{SO(10)}).
In the SO(10) case, the flavor violation pattern in the lepton sector
and the quark sector can depend on the SO(10) symmetry breaking vacua.
Actually, in order to forbid a rapid proton decay,
the quark flavor violation should be larger than the lepton flavor violation
among the symmetry breaking vacua \cite{Dutta:2007ai}.
Namely, it is expected that
$\kappa_{Q}$, $\kappa_{U^c}$, and $\kappa_{D^c}$ are much larger than
$\kappa_L$ and $\kappa_{E^c}$.
For example, if only the Higgs fields $({\bf 8},{\bf 2}, \pm1/2)$
are light compared to the breaking scale (which is the most suitable case),
one obtains $\kappa_Q = \kappa_{U^c} = \kappa_{D^c}$,
and only quark flavor violation is generated, while the lepton flavor
violation is not generated.
On the other hand, when the flavor violation is generated
from the minimal type of SU(5) vacua with the type I seesaw,
it is expected that $\kappa_{L}$ is always larger than $\kappa_{D^c}$
since the right-handed Majorana mass scale is less than the scale of
colored Higgs mass.
Therefore, the existence of $b$-$s$ transition indicated by the
experimental results in Fermilab
predicts the sizable lepton flavor violation in the minimal type of
SU(5) model.
In other words, if the results of a large $B_s$-$\bar B_s$ phase is really
an evidence of NP,
the minimal type of SU(5) GUT models are restricted
severely \cite{Parry:2005fp,Dutta:2008xg,Hisano:2008df}.

\section{Constraints from the neutralino dark matter}

The cosmic microwave background anisotropy measurement by WMAP~\cite{WMAP} put a stringent constraint on the SUSY parameter space through the dark matter requirement. Within 2~$\sigma$, the neutralino relic density should be
$0.106 < \Omega h^2 < 0.121$.
This assuming that dark matter consists solely of neutralino, i.e. smaller relic density cannot be a priori excluded.  In SUSY models with universal gaugino and sfermion masses, $m_{1/2}$ and $m_0$ respectively, it is well known that there are five distinct regions that satisfy the relic density constraint: (a) the bulk region
where both $m_{1/2}$ and $m_0$ are small, (b) the neutralino-stau coannihilation region where the lightest stau mass is almost degenerate with the neutralino mass, (c) the focus point (FP)/hyperbolic region at large $m_0$ where the $\mu$ parameter becomes small and the lightest neutralino gets more Higgsino content, (d) the funnel region where the heavy Higgs masses ($m_A$ and $m_H$) is about twice the neutralino mass, and (e) the neutralino-stop coannihilation region where the lightest stop mass is suppressed by large off-diagonal terms when the trilinear coupling parameter $A_0$ is large. In our analysis we assume that the Higgs soft masses are not tied to $m_0$, and from this assumption we have two more free parameters $\mu$ and $m_A$. In such models there could be another dark matter region, in addition to the five regions above, i.e. the neutralino-sneutrino coannihilation region at large $m_A$ and/or $\mu$~\cite{sneutrinoNLSP}.

The neutralino dark matter hypothesis is very attractive, and there are lots of activities, experimentally and theoretically, to discover the dark matter candidate. As a weakly interacting massive particle (WIMP), the lightest neutralino can in principle be detected directly by ultra-sensitive detectors. Such experiments are now reaching $O(10^{-8}~{\rm pb})$ sensitivity level for certain values of neutralino mass~\cite{CDMSII,Xenon100}. Neutralino particles in the galaxy can also give out some indirect signals through their annihilation, in particular from some regions where neutralinos can accumulate due to some gravitational potential attractors. It was pointed out that high energy neutrino flux from the sun can potentially be a clear signal of dark matter annihilation in the sun~\cite{Silk:1985ax}, and this is currently being searched by the IceCube experiment and its upgrade with DeepCore which can lower the neutrino energy threshold for the detector~\cite{IceCube}.

To study the dark matter aspect of our model, we calculate the neutralino relic density, the neutralino-proton elastic scattering cross section, and the muon flux induced by solar neutrinos from neutralino annihilation. For the muon flux calculation we use the {\tt DarkSUSY} program version 5.0.5 \cite{DarkSUSY} which utilizes the results of {\tt WimpSim}~\cite{wimpsim}, interfaced with our own spectrum program. Solar neutrino from WIMP dark matter in various models has recently also been analyzed in~\cite{EOSS,BKMS,solnu}. We assume the NFW profile~\cite{Navarro:1995iw} for our analysis. As mentioned in \cite{EOSS}, there are uncertainties in the neutralino-nucleon cross section due to the strange quark role in the interaction. We use their default values for $\Sigma_{\pi N}$ and $\Delta_s$, i.e. $\Sigma_{\pi N} = 64$~MeV, and $\Delta_s^{(p)} = -0.09$.

Since the parameter space in the minimal type of SU(5) is restrictive,
(in other words, the mass spectrum is constrained),
 the discovery potential of this region
appears to be very promising at the LHC,
and at the direct and indirect dark matter search experiments~\cite{Dutta:2009hj}.
Since small values of  $\mu$ are not preferable
due to the $\tau\to\mu\gamma$ constraint,
the WMAP relic density
prefers the funnel solution
(i.e. the neutralinos annihilate through the heavy Higgs bosons pole).
It is also true
that a small value of $m_A$ is preferred to enhance
the double penguin contribution.
In that case, the
spin-independent neutralino-nucleon scattering cross section
can be enhanced.

\begin{figure}[tbp]
 \center
 \includegraphics[viewport = 10 0 220 200,width=8cm]{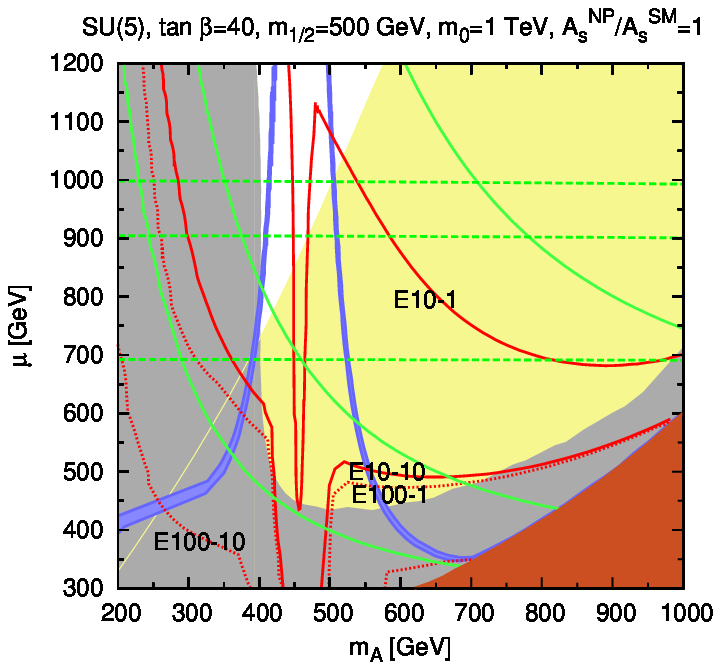}
 \includegraphics[viewport = 10 0 220 220,width=8cm]{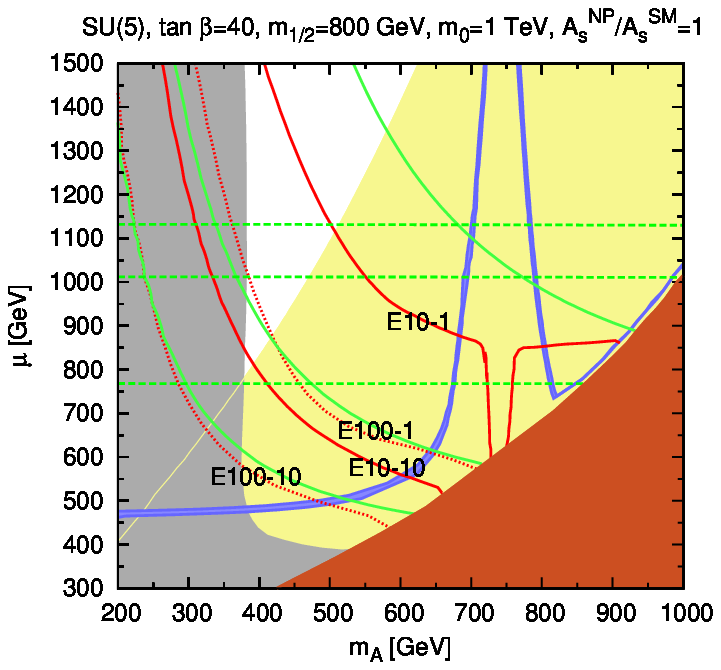}
 \caption{(Left) The $m_A - \mu$ plane in SU(5) model with $\tan \beta = 40$, $m_{1/2} = 500$~GeV, $m_0 = 1$~TeV, $A_0 = 0$, and $A_s^{\rm NP}/A_s^{\rm SM} = 1$, showing various constraints from flavor and dark matter sectors as discussed in the text. (Right) Same plot for $m_{1/2} = 800$~GeV.
}\label{fig:mumA}
\end{figure}

In figure~\ref{fig:mumA},
we show the allowed region in the $m_A$-$\mu$ plane
when $A_s^{\rm NP}/A_s^{\rm SM} = 1$
in the SU(5) case, $\kappa_L = \kappa_{D^c}$.
For the left panel we choose as SUSY parameters $\tan \beta = 40$, $m_{1/2} = 500$~GeV, $m_0 = 1$~TeV and $A_0 = 0$. The same parameters are used for the right panel except for $m_{1/2} = 800$~GeV.
The yellow and gray regions are excluded by the $\tau\to\mu\gamma$
and $B_s\to\mu\mu$ constraints, respectively.
The red-brown region is excluded because the lightest stau is the LSP there, and therefore the neutralino cannot be the dark matter candidate. The WMAP relic density range is obeyed for the neutralino in the narrow blue bands. The solid green lines are contours for the scalar neutralino-proton elastic scattering cross section of $5, 1, 0.1 \times 10^{-8}$~pb respectively from left to right. The almost horizontal dashed green lines are for the spin-dependent cross section, $5, 10, 50 \times 10^{-8}$~pb
respectively from top to bottom. We also show contours of muon flux, labeled as Ex-y, where x is the assumed detector energy threshold in GeV and y is the flux in km$^{-2}$yr$^{-1}$.  We show two cases for E(threshold): 100 and 10~GeV, dotted and solid respectively. As we can see, the Br($\tau\to\mu\gamma$) constraint is important. In the left plot, the funnel happens at relatively small $m_A$, and a large part is allowed. In the right plot, however, the funnel is shifted to the right due to the larger neutralino mass. In the later case, we need much larger $\mu$ to satisfy the Br($\tau\to\mu\gamma$) constraint, although there would be an upper bound on $\mu$ due to the sneutrino LSP region~\cite{sneutrinoNLSP}.

The muon flux rate is correlated to the neutralino-proton scattering cross section since this cross section determines the number of neutralinos accumulated in the core of the sun, hence the neutralino annihilation rate there. Since protons constitute a large portion of the sun, both the spin dependent and independent neutralino-proton cross sections are comparably important. For most of the MSSM parameter space the spin dependent part is larger than the scalar part, and this leads to a widespread misconception that for solar neutrino flux calculation only the spin dependent cross section is important. However, there are some regions of the parameter space where the scalar and the spin dependent cross section are about the same order of magnitude, and in this case the scalar contribution is also significant in determining the flux. This was also pointed out by \cite{EOSS}. We can see in Fig.~\ref{fig:mumA} that for small $m_A$ the scalar cross section is quite large, and the
 muon flux is following the scalar contour. For large $m_A$, however, the spin dependent part becomes larger than the scalar part, and the muon flux contour is flatten out. Furthermore, the solar neutrino flux also depends on the branching fractions of the neutralino annihilation. Near the stau LSP region, the lighter stau mass is relatively small, enhancing the $\tau^+ \tau^-$ channel, which in turn increases the flux (visible on the left panel).
In the middle of the funnel region, the neutralino relic density is much below the lower bound of the WMAP range. We rescale down the muon flux due to this fact, and this is seen as a drops on the muon flux rate contour.

The IceCube with DeepCore can potentially detect neutrinos with energy down to 10~GeV~\cite{jenni}. It appears that for the left plot large allowed region, i.e., the entire left band of the funnel region and part of the right band, can be detected. Note, however, that we should also consider the backgrounds to have a clear discovery~\cite{BKMS}, and therefore sufficiently large amount of data would need to be collected. For the plot on the right side, however, it would be very difficult for the IceCube with DeepCore to probe the WMAP region not already excluded by the Br($\tau\to\mu\gamma$) constraint.

In the SO(10) case, the figures for the same parameter space remain unchanged qualitatively,
except for a lower $\mu$ region where
the chargino contribution of the Higgs penguin
from the left-handed squark FCNC is important.
As mentioned in the previous section,
when the SO(10) breaking vacua is chosen to satisfy the
proton decay constraint while gauge unifications are maintained,
there is no stringent constraint from $\tau\to\mu\gamma$,
and a larger region of parameter space is allowed.
Consequently, the dark matter direct and indirect detection will play more significant roles in excluding the parameter space.

\section{Discussions}

In this paper, we investigated the effect of the recent dimuon CP asymmetry from $B$ decay modes
observed at 3.2 $\sigma$ deviation from the Standard Model by
the D0 collaboration in the context of $R$-parity conserving SUSY GUT models
and show that a large amount of flavor violation between the second
and the third generation can be generated.
Not only the large flavor violation arises due to large atmospheric mixing angle,
but also new CP phases are obtained from the Yukawa interactions
in grand unification,
and they are responsible for the observed large CP asymmetry.

Because of the quark-lepton unification,
the CP asymmetry is restricted due to the $\tau\to\mu\gamma$ bound.
This  restriction depends on the
source of flavor violation (Dirac neutrino Yukawa coupling
in the minimal type of SU(5)
 or Majorana neutrino Yukawa coupling in the minimal type of SO(10)),
SUSY mass spectrum (larger SUSY breaking masses are preferred),
and dominant diagram (box diagram for lower $\tan\beta$
or double Higgs penguin diagram for larger $\tan\beta$).
We find that large values of $\tan\beta$ ($=30-50$)  are preferred
because the CP asymmetry is enhanced via the
double Higgs penguin diagram, whose contribution is
proportional to $\tan^4\beta$
and a sizable contribution to the flavor violating $b$-$s$ transition can be
easily realized.
We found that
the minimal type of SO(10) is preferred
due to the fact that both left- and right-handed
squark mass matrices can have FCNC sources.
The intermediate values of $\tan\beta$ ($\tan\beta = 20-30$)
are not very preferable.
In the case of the minimal type of SU(5),
$\tan\beta$ should be large to make the double penguin diagram
dominant.
In that case, $B_s\to\mu\mu$ is enhanced and
it provides a lower bound of Br($B_s\to\mu\mu$),
which is about $1\times 10^{-8}$,
in order to obtain a large CP asymmetry.

The symmetry breaking vacua can be chosen to
make the quark-lepton unification relaxed in the SO(10) case,
while in the SU(5) case where the neutrino Dirac Yukawa coupling is
the source of FCNCs, the leptonic FCNC is always larger
than the quark FCNC.
Therefore, the bound from the LFV in the SU(5) is more stringent,
and in other words, the spectrum is more predictive
to obtain the large CP asymmetry indicated by the like-sign
dimuon charge asymmetry.
In fact, in order to satisfy the dark matter content of the universe, the CP asymmetry prefers the funnel solution
where the lightest neutralinos annihilate through the heavier Higgs bosons pole.
Such a restriction on the spectrum from the flavor physics
would allow us to observe this parameter space at the LHC and at the direct and indirect
dark matter detection facilities more easily.
We showed that the high energy neutrino flux from the sun
from the neutralino annihilation can be detectable at IceCube with DeepCore for some regions of parameter space.

\section*{Acknowledgments}

We thank W.S. Hou, X.G. He, S. Khalil, S. Su and X. Tata
for useful discussions. Y.S. thanks J.~Edsjo, D.~Marfatia, K.A.~Richardson-McDaniel and E.~Sessolo for various information regarding solar neutrino analysis. B.D. thanks the hospitality of the GGI, Florence, where part of the work was done.
The work of B.D. is supported in part by the DOE grant DE-FG02-95ER40917.
The work of Y.S. is supported in part by the DOE grant DE-FG02-04ER41308.
The work of Y.M. is supported by the Excellent Research Projects of
National Taiwan University under grant number NTU-98R0526.

\end{document}